# Parametric Resonance May Explain Virologic Failure to HIV Treatment Interruptions

Romulus Breban and Sally Blower[*]

*Department of Biomathematics and UCLA AIDS Institute, David Geffen School of Medicine, University of California, Los Angeles, California 90024*

[*] To whom correspondence should be addressed. E-mail: sblower@mednet.ucla.edu

**Pilot studies of structured treatment interruptions (STI) in HIV therapy have shown that patients can maintain low viral loads whilst benefiting from reduced treatment toxicity. However, a recent STI clinical trial reported a high degree of virologic failure. Here we present a novel hypothesis that could explain virologic failure to STI and provides new insights of great clinical relevance. We analyze a classic mathematical model of HIV within-host viral dynamics and find that nonlinear parametric resonance occurs when STI are added to the model; resonance is observed as virologic failure. We use the model to simulate clinical trial data and to calculate patient-specific resonant spectra. We gain two important insights. Firstly, within an STI trial, we determine that patients who begin with similar viral loads can be expected to show extremely different virologic responses as a result of resonance. Thus, high heterogeneity of patient response within a STI clinical trial is to be expected. Secondly and more importantly, we determine that virologic failure is not simply due to STI or patient characteristics; rather it is the result of a complex dynamic interaction between STI and patient viral dynamics. Hence, our analyses demonstrate that no *universal* regimen with periodic interruptions will be effective for all patients. On the basis of our results, we suggest that immunologic and virologic parameters should be used to design patient-specific STI regimens.**

**Introduction**

STI [1-6] are therapeutically prescribed interruptions of HIV continuous therapy. There are three major advantages to STI. First, STI decrease toxicities and side effects of HIV treatments [7, 8]. Second, STI decrease the cost of HIV therapy, this being a particularly important consideration for resource-limited settings. Third, STI may stimulate HIV specific immune responses (*autoimmunization*), thus patients may gain therapeutic benefits that are inaccessible through continuous therapy [5, 6, 9, 10]. Patients entering STI trials are typically selected on the basis of their viral load and CD4 count. Interruption and resumption of therapy can be dictated by viral load and CD4 cell count [6, 11], can be periodically scheduled, or can have other protocols [12]. Periodic interruption schedules including 5 days on/2 days off [13], one week on/one week off [7, 8, 14], 3 weeks on/one week off [15], 8 weeks on/2 weeks off [16], 8 weeks on/4 weeks off [17], 30 days on/30 days off [18], 6 months on/one month off [10] have been tried for various purposes and with various degrees of success. The most intensively studied periodic STI is one week on/one week off, specifically designed for maintaining low viral load while decreasing toxicities and cost. The two pilot studies [7, 8] of this treatment schedule showed zero virologic failure. However, the only large-scale clinical trial [14] of this treatment schedule was prematurely terminated due to a virologic failure rate of 53%. Currently, the reason for this high rate of virologic failure is unknown; drug resistance does not appear to be the cause [14]. With the claim to have learned from the experience of one week on/one week off STI trials, a new STI pilot study of 5 days on/2 days off awaits scaled-up investigations [13]. Here we present a new hypothesis that may explain virologic failure during regimens with periodic treatment interruptions. We analyze a mathematical model to show that STI can interact with viral dynamics and cause (nonlinear) resonance; this interaction can be observed as virologic failure.

Resonance is a general phenomenon that has long been recognized in physics and engineering [19-21]. Recently, resonance has been proposed as a mechanism in population biology to explain the seasonality of the influenza epidemics [22]. We hypothesize that resonance is to be expected at virologic level, as well. Consider a system at equilibrium (e.g., viral set point) that has its own natural period. In the presence of an external periodic perturbation (e.g., a periodic change in parameters), the system tends to follow the perturbation and displays oscillations in the dynamic variables. The phenomenon of resonance represents the selectivity of the system to the external periodic perturbation. If the period of the external perturbation is close to a critical value (determined by the natural period of the system), resonance will occur. At resonance, the system responds strongly to the external perturbation and therefore the dynamic variables fluctuate widely, whilst when the system is far from resonance the fluctuations are much reduced. Resonance is called *parametric* if the perturbation applied to the system is a periodic change in parameters. Resonance is also classified as linear and nonlinear. In the case of small periodic perturbations (e.g., STI of ineffective treatment), we would expect linear parametric resonance (see supplemental material). In the case of large periodic perturbations (e.g., STI of effective treatment), we would expect nonlinear parametric resonance.

The last decade of research in virology has demonstrated that mathematical models are extremely useful tools for understanding viral dynamics. Simple within-host models [23-32] have been developed and successfully used to provide greater understanding of HIV, hepatitis C, and hepatitis B. Treatment has been modeled by changing the effective values of the infection parameters. However, the interaction between treatment and virus can also have dynamical aspects. Simple models of viral dynamics have revealed that the viral set point (i.e., within-host endemic equilibrium) is reached through strongly damped oscillations [23, 25, 32]. In practice, viral oscillations have not been observed. This may be due to strong damping, coarse data sampling, or accuracy limitations of the measurements. The damped oscillations around the viral set point have a well-defined period (i.e., duration of an oscillation cycle), and thus, a system is formed that is potentially subject to resonance. A regimen with periodic treatment interruptions acts as an external perturbation. If the STI period coincides with the critical resonance period of the viral dynamics, large fluctuations in the viral load will occur. To investigate our resonance hypothesis, we analyzed a classic model of HIV within-host viral dynamics [23, 25, 27, 29] with added STI (see methods), and we simulated clinical trial data. Since resonance is a model-independent phenomenon, analysis of a simple HIV model is adequate for obtaining general, qualitative results on how resonance manifests in viral dynamics.

**Results and Discussion**

We first used this model to simulate the virologic response of two patients [A (red) and B (blue)] participating in a theoretical STI clinical trial; Fig. 1(a). The immunologic and virologic parameters used are given in Table 1. Since current treatments to not cure HIV, we must choose the treatment parameters such that the reproduction ratio of the HIV infection under treatment is reduced but still larger than 1. The two patients A and B become infected with HIV at time $t = 0$, reach viral set points through strongly damped oscillations, and then enter the STI trial. Only the first and largest peak in the viral load may be observable and corresponds to the primary infection response. Subsequent oscillations are heavily damped, additional peaks might be very hard to observe in practice as they are very close to the viral set point; see Fig. 1. Note that the viral loads of A and B appear indistinguishable at trial entry; thus, it may be expected that these two patients should respond very similarly to the STI. Surprisingly, the viral load peaks of A under the STI regimen are approximately twice as high as those of B. This occurs because although the set points of A and B are very similar, some of the parameters that determine the set point are very different (Table 1). These parameters play a crucial role for the natural oscillation about the viral set point; hence, the STI resonates with the viral dynamics of A, but not with that of B. Due to resonance selectivity, patients who appear indistinguishable at entry to a clinical trial can respond very differently. In contrast, we find that patients A and C [see Table 1 and Fig. 1(b)], who have very different pre-STI set points, and hence would be expected to respond differently, actually respond similarly to the STI. This is also a consequence of resonance selectivity; in this case the STI does not resonate with the within-host dynamics of C, and therefore C displays only small fluctuations in the viral load that are comparable to those of A. We have shown that patient response to an STI regimen is heterogeneous and difficult to predict. Blood tests measuring the viral load (and the CD4 count) before entering an STI regimen provide insufficient information to anticipate how well a patient will perform under the STI regimen.

To investigate the effect of the STI period (i.e., duration of an interruption cycle) on virologic failure, we simulated a second STI clinical trial. This provided the patients with the same number of days on treatment as the first STI, only with a different period; see Fig. 1(c). The periodicity of the STI in Trial 1 resonated with the within-host dynamics of A but not with that of B. The periodicity of the STI in Trial 2 did not resonate with the within-host dynamics of either A or B. Therefore, in Trial 2 (in contrast to Trial 1) the virologic responses of A and B were similar, as demonstrated by Fig. 1(c). To understand possible virologic responses of A, B, C to a large range of periodicities of STI, we constructed patient-specific resonance spectra. These spectra show the maximum (thin line), minimum (thick line), and average (dashed line) virologic response for each patient as a function of the STI periodicity; see Fig. 2. It may be seen that the maximum viral load varies significantly both with the STI periodicity and from patient to patient. For example, patient C would fail (i.e., have high fluctuations in the viral load) an STI with short periodicity, whilst patient A would fail an STI with large periodicity. Also, patients A and B, although having similar pre-treatment viral loads, display very different resonance profiles.

We hypothesize that the increased heterogeneity that has been observed in large STI clinical trials may be explained by recognizing that within-host viral dynamics of patients may resonate with the STI, and that patients may have a wide variety of resonance profiles. The heterogeneity of immunologic and virologic parameters of patients is likely to increase with trial size; hence increased virologic failure rates are to be expected in large trials. Only patients not having resonant interactions with the STI regimen would respond well, but this may be a relatively small fraction of the patients.

By using a classic mathematical model of within-host HIV dynamics we have shown that a nonlinear resonant interaction can occur between within-host viral dynamics and STI. Our hypothesis provides a new understanding of virologic failure during STI. We have demonstrated that patients with similar pre-STI viral loads may or may not respond in the same way to STI, depending on the STI. We have also shown that patients with high pre-STI viral loads do not necessarily respond worse than those with low pre-STI viral loads. The interaction between the within-host viral dynamics of a patient and STI is characterized by a resonance profile which may vary significantly from patient to patient. Based on the idea of resonant interaction, great heterogeneity of patient response is likely to occur in large STI clinical trials. Thus, as a consequence of our results, no STI may be expected to be successful for a large number of patients. Hence, we suggest that STI regimens should be tailored to patient-specific immunologic and virologic parameters.

**Methods**

We analyzed a classic mathematical model of the HIV infection that has been developed by others [23, 25, 27, 29], and has also been used for the study of hepatitis B [26, 31] and hepatitis C [24], in order to determine whether resonance occurs when STI are added to this model. For the HIV viral dynamics, this model gives results qualitatively similar to other models [32] which are also potentially subject to resonance. More sophisticated models containing CTL dynamics [30] can be studied on similar mathematical grounds. The model equations are

$$dx/dt = \lambda - dx - \beta xv,$$
$$dy/dt = \beta xv - ay,$$
$$dv/dt = ky - uv,$$

where $x$ denotes the population of uninfected CD4 lymphocytes, $y$ denotes the population of infected CD4 lymphocytes, and $v$ denotes the virus population. The parameters of the system are as follows. $\lambda$ is the birth rate of the CD4 lymphocytes. $\delta$, $a$, and $u$ are the death rates of the uninfected CD4 lymphocytes, infected CD4 lymphocytes, and virus, respectively. $\beta$ is the infectiousness of the virus -- i.e., the fraction of the possible contacts between CD4 cells and virus that yield infected CD4 cells -- and $k$ is the rate at which an infected cell produces new virus particles. The parameter values are taken from [32], and they are listed in Table 1.

We modeled the effect of treatment as a decrease by 30% in the virus infectiousness (i.e., $\beta$) and production rate (i.e., $k$) of the infection. Even under treatment, the parameters yield a basic reproduction ratio which is larger than 1; this reflects the real world fact that current treatment does not cure HIV. We did not include in the model the pharmacokinetics of the drug concentration in the blood and its influence on the infection parameters (see for example [32]) since this develops on the time scale of 6 to 8 hours and interferes very little with the much slower viral dynamics on the time scale of weeks. At the time scale of weeks, an average of the rapid fluctuations of the infection parameters influenced by the drug represents a good approximation of the exact variation driven by the pharmacokinetics.

Designing STI, we choose that the amount of time a patient stays on treatment is equal to the amount of time off treatment but this does not need to be so for a resonant interaction to occur. Modeling other periodic interruption regimens can be approached in a similar fashion. We discuss drug naïve patients entering STI trials, but similar qualitative results are obtained for drug experienced patients, as well.

Our hypothesis of resonance can be tested in clinical studies of periodic treatment interruptions. Given a group of HIV patients that have been carefully measured for their immunologic and virologic parameters (see e.g., [28]), an experiment can be designed in two basic ways. First, selecting HIV patients with similar infection parameters, each of these patients is set on the same drug regimen but with different periodic interruptions. Then, close monitoring of the viral load would enable for the computation of resonance profiles like those in Fig. 2. Second, selecting HIV patients with significantly different infection parameters, each of the patients is set on the same drug regimen with the same periodic interruptions. Since the patients have different infection parameters, the STI may induce resonance in some but not all of the patients. In this case, close monitoring of the viral load would reveal that large fluctuations in the viral load occur in selected patients with certain infection parameters, demonstrating resonance as in Figs. 1(a,b). Most importantly, such an experiment would certify that there exists no universal regimen with periodic interruptions applying to *all* patients. Rather, regimens with periodic interruptions must be tailored to specific immunologic and virologic parameters of the patient.

**Supplemental material** accompanies the paper on the AIDS web site.

**Acknowledgements.** SB and RB gratefully acknowledge the support of NIH/NIAID RO1 AI041935.

**Competing interests statement.** None.

**Figure captions**

**Figure 1** Simulations of viral loads of patients versus time. (a) Patients A (red) and B (blue), infected at time $t=0$, undergo no treatment for one year. The viral loads settle to viral set points through damped oscillations. After that, the patients enter an STI regimen with 80 days on treatment and 80 days off treatment (160 day period). The shaded areas represent the intervals of time on treatment; (b) the same as (a) for patients A (red) and C (green); (c) the same as (a) but with 125 days on treatment and 125 days off treatment (250 day period).

**Figure 2** The simulated maximum (thin line), minimum (thick line), and average (dashed line) of the oscillations in the viral load versus the period of STI for (a) patient A, (b) patient B, and (c) patient C. The positions of the STI regimens in Fig. 1 are marked with dotted lines. The left dotted line corresponds to the STI regimen of Figs. 1(a,b), and the right dotted line corresponds to the STI regimen of Fig. 1(c).

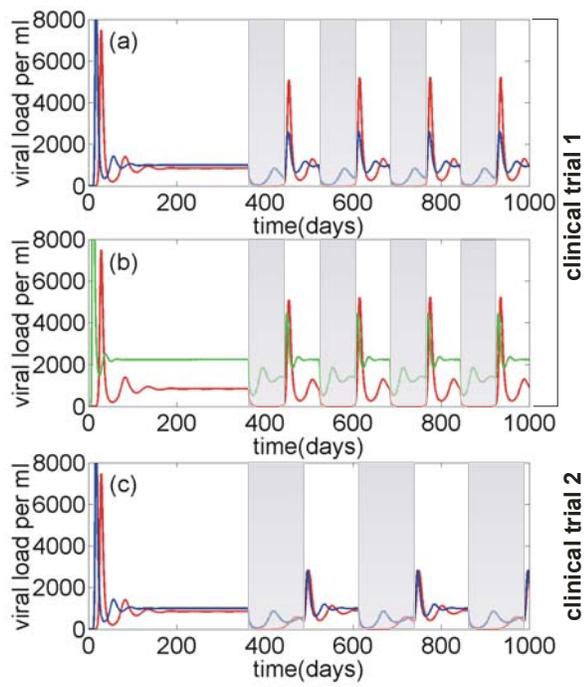

Fig. 1

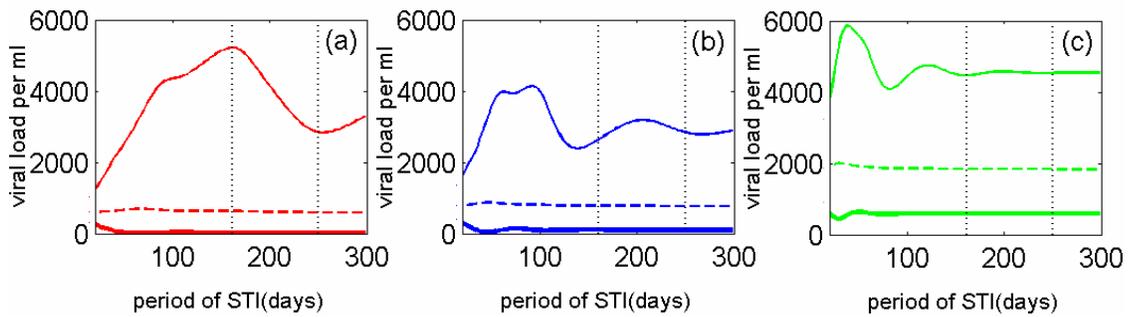

Fig. 2
14

**Table 1. Parameter values.** Parameter estimates of the HIV model for patient A [32]. Patient B has the same parameters as patient A with the exception of $\beta$ which $\beta = 8 \times 10^{-5}$/(cell/ml)/day. Patient C has the same parameters as patient A except $k$ which is $k = 200$/day/ml. We have obtained qualitatively similar results using other parameter sets provided in the literature [27]. The effect of treatment is modeled as a decrease by 30% in $\beta$ and $k$.

| Parameter | Biological Interpretation | Unit | Value | | |
|---|---|---|---|---|---|
| | | | Patient A | Patient B | Patient C |
| $\lambda$ | Immigration rate of CD4 cells | cells/day/ml | 20 | 20 | 20 |
| $d$ | Uninfected CD4 cell death rate | 1/day/ml | 0.02 | 0.02 | 0.02 |
| $a$ | Infected CD4 cell death rate | 1/day/ml | 0.4 | 0.4 | 0.4 |
| $u$ | Free virion death rate | 1/day/ml | 4.0 | 4.0 | 4.0 |
| $\beta$ | Infectivity parameter | 1/(cell/ml)/day | $5 \times 10^{-5}$ | $8 \times 10^{-5}$ | $5 \times 10^{-5}$ |
| $k$ | Virus production rate | 1/day/ml | 100 | 100 | 200 |